\documentclass[%
reprint,
superscriptaddress,
 amsmath,amssymb,
 prl, 
showpacs
]{revtex4-2}
\usepackage{float}
\usepackage{graphicx}
\graphicspath{./Figures}
\usepackage{dcolumn}
\usepackage{bm}
\usepackage[caption=false]{subfig}
\usepackage[export]{adjustbox}
\usepackage{mathptmx} 
\usepackage{mathtools}

\begin{document}
\preprint{APS/123-QED}
\title{Simultaneous 2D and 3D turbulent flows in Faraday
Waves}

\author{Raffaele Colombi}
 \email{raffaele.colombi@tuhh.de}
\homepage{\\https://www.tuhh.de/ims/research/2d-turbulence-in-faraday-flows.html}
\author{Niclas Rohde}%
\author{Michael Schl\"uter}
\affiliation{%
 Institute of Multiphase Flows, Hamburg University of Technology
}%


\author{Alexandra von Kameke}
\email{alexandra.vonkameke@haw-hamburg.de}
\homepage{\\https://www.haw-hamburg.de/hochschule/beschaeftigte/\\detail/person/person/show/alexandra-kameke/172/}
\affiliation{
 Department of Mechanical Engineering and Production Management, 
Hamburg University of Applied Sciences 
}%

\date{\today}

\begin{abstract}
In nature turbulent flows exist that are neither simply 2D nor 3D but boundary conditions, such as varying stratification, force them towards the one or the other. Here, we report the first evidence of the co-existence of 2D and 3D turbulence in an experimental flow driven by Faraday waves in water. We find that an inverse energy cascade at the fluid surface and a direct energy cascade in the 3D bulk flow underneath exist simultaneously. We base our analysis on temporally and spatially well-resolved velocity fields measured at horizontal and vertical planes. The findings suggest that the strongly turbulent 2D surface flow drives the 3D bulk flow through sporadic vertical jets as a source of momentum. 
%
\end{abstract}


\maketitle

\begin{center} 
\begin{figure}[H] 
\setlength{\abovecaptionskip}{10pt}
\setlength{\belowcaptionskip}{-40pt} 
\setlength\intextsep{3pt}
    \subfloat[]{%
      \includegraphics[height=5.4cm,valign=b]{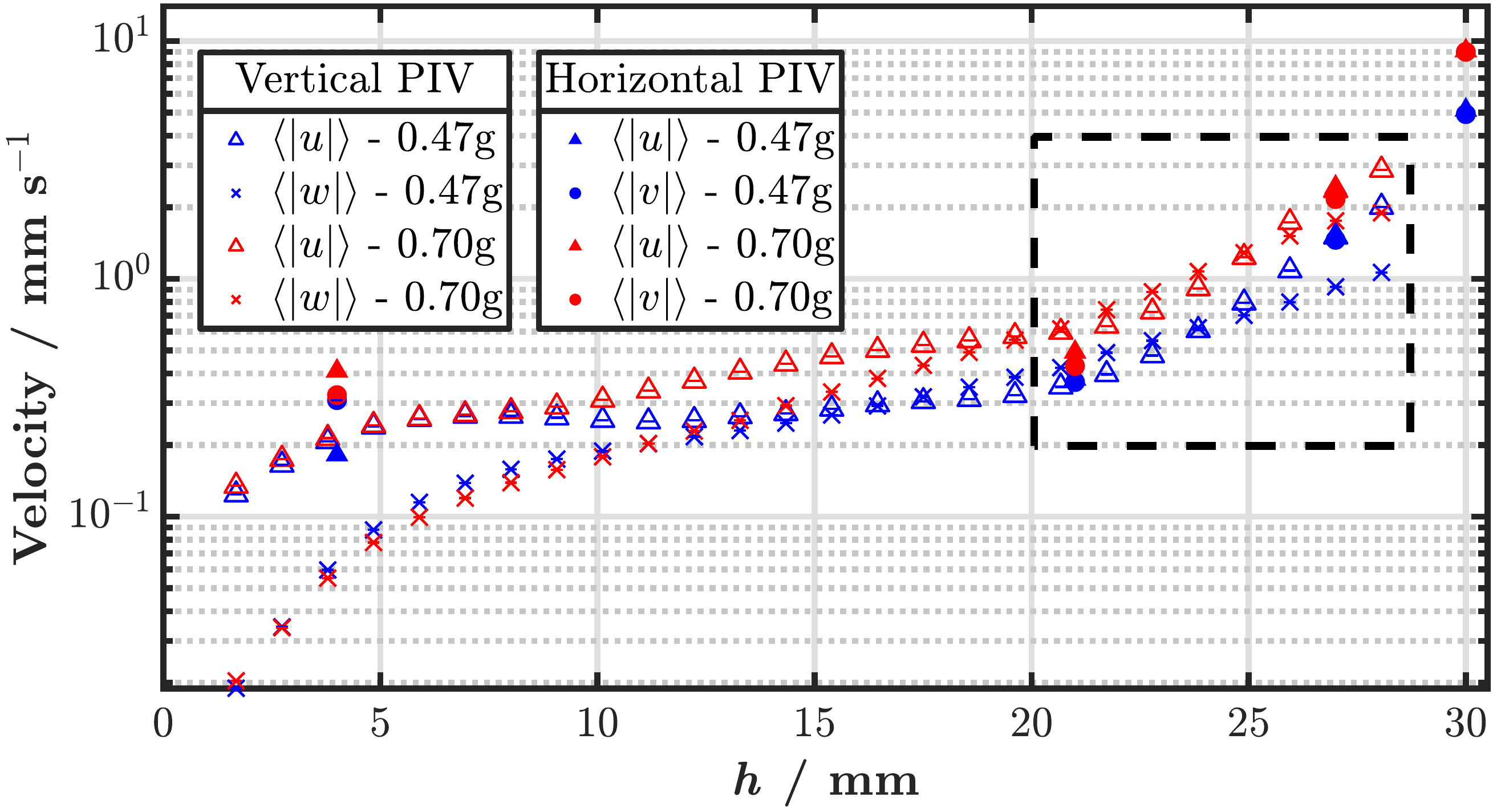}
    }
    \hfill
    \subfloat[]{%
      \includegraphics[height=5.4cm, valign=b]{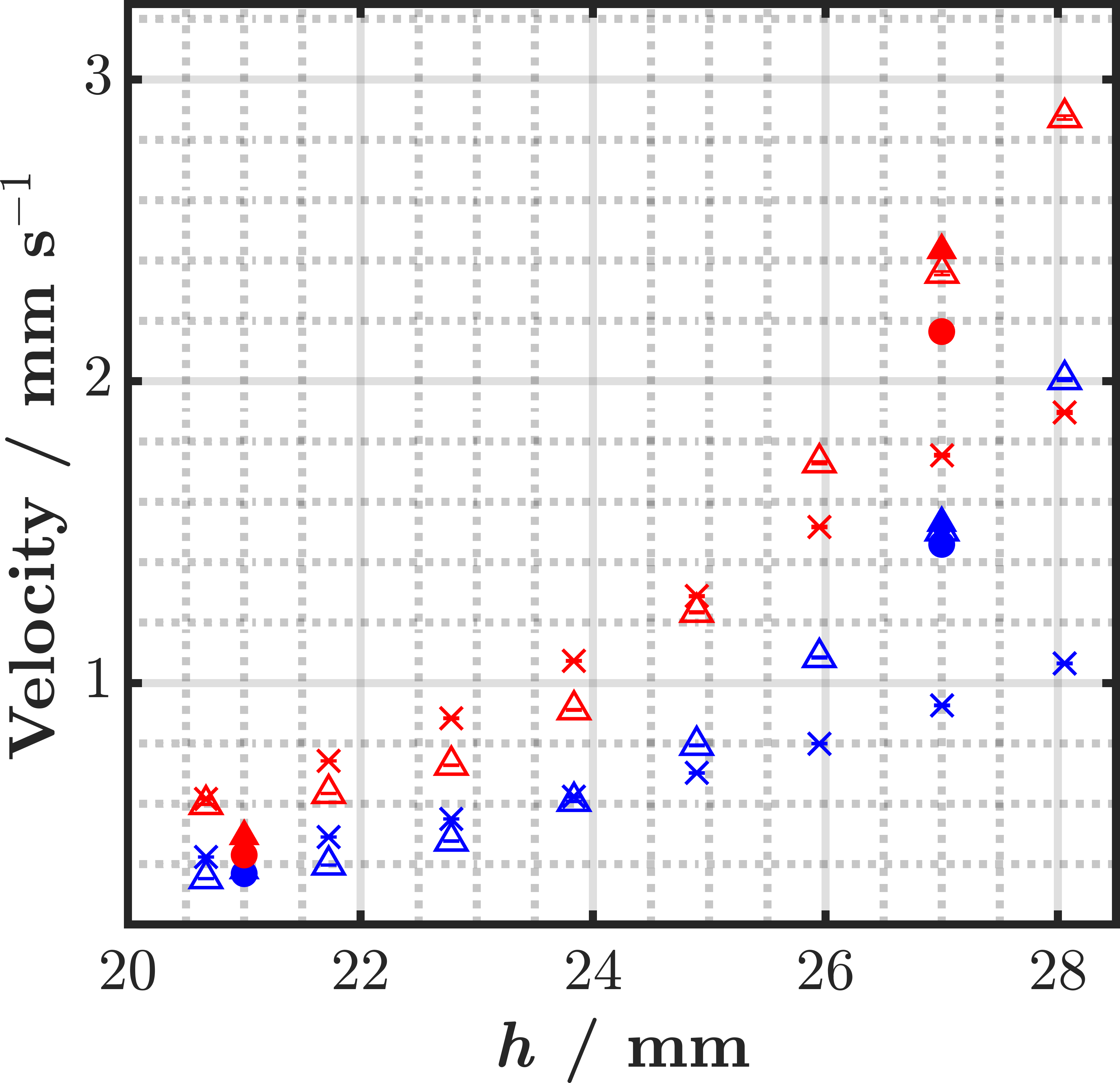}
    }
\hspace{0.5cm}
\caption{a) Profiles of absolute values of velocity components \( \langle|u|\rangle\), \( \langle|v|\rangle\), and \( \langle|w|\rangle\) against the distance from the container bottom \(h\) for vertical and horizontal PIV measurements (empty and filled markers respectively) and forcing accelerations \(a_f = 0.47\)g and \(a_f = 0.70\)g (blue and red  markers respectively). Values are averaged across available time-steps, measurement runs (4 or 6, \citep{supplementalMaterial}) and PIV grid points. b) Zoom-in region in the layer beneath the fluid surface (thickness of one Faraday wavelength). Color online.}
\label{fig:vel_profiles}
\end{figure}
\end{center}

\begin{center}
\begin{figure}[H]
\setlength{\belowcaptionskip}{-32pt}
    \subfloat[$h=30$ mm (Surface)]{%
      \includegraphics[height=7.3cm]{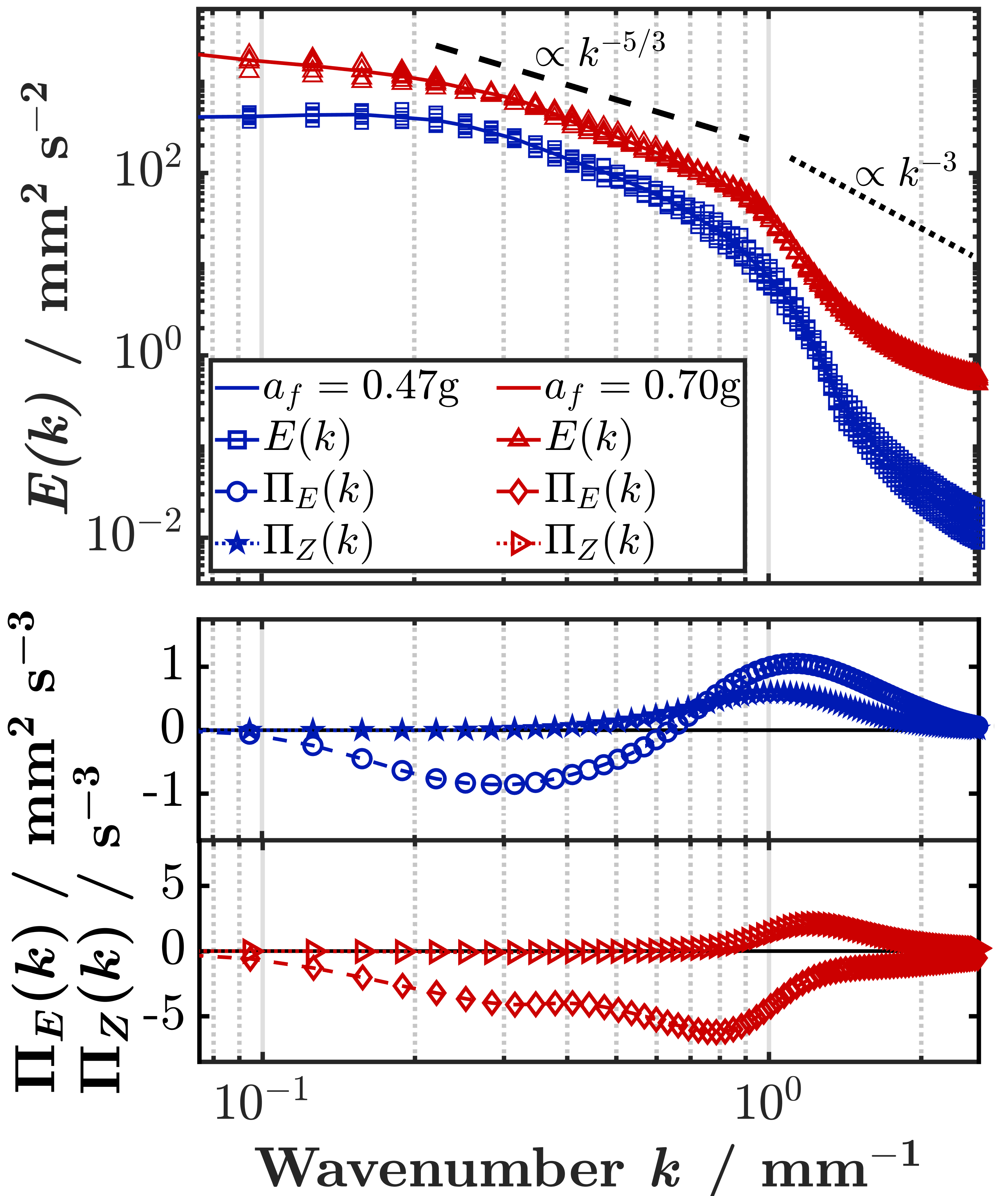}
    }
    \hfill
    \subfloat[$h=27$ mm]{%
      \includegraphics[height=7.3cm]{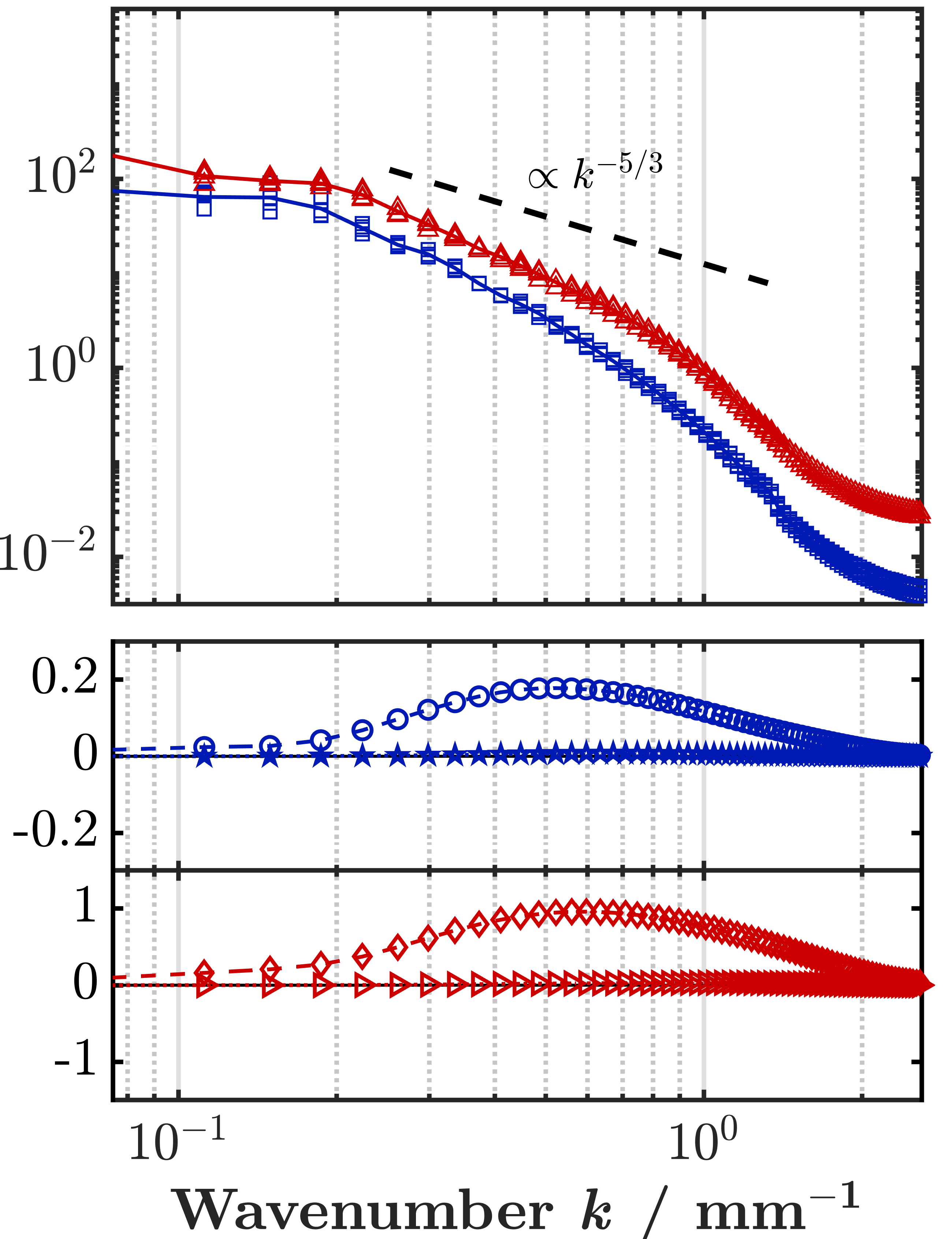}
    }
    \hfill
    \subfloat[$h=21$ mm]{%
      \includegraphics[height=7.3cm]{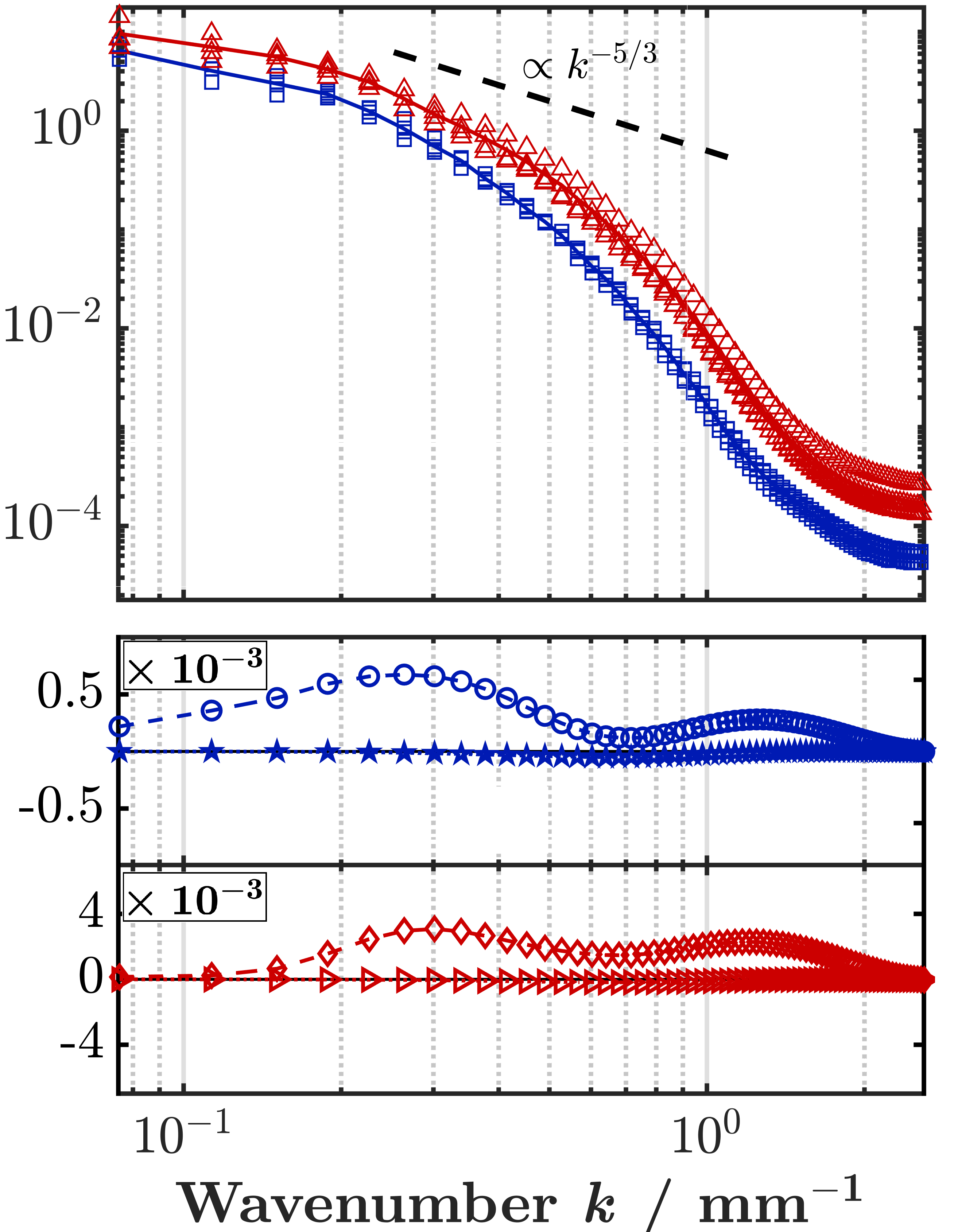}
    }
    \caption{Wavenumber spectrum of flow kinetic energy, net energy fluxes and net enstrophy fluxes at different horizontal planes for forcing accelerations \(a_f=0.47\)g and \(0.70\)g (blue and red). Results are averaged for all available time steps and measurement runs (4 or 6, \citep{supplementalMaterial}). Color online.}
    \label{fig:energy_spectra}
 \end{figure}
 \end{center}

The study of two-dimensional three-component (2D3C) flows has placed particular emphasis on the occurrence of an inverse energy cascade in thick layers and the transition from 2D to 3D turbulence and vice versa \citep{biferale2017two, kokot2017active}. Examples of experimental flows exhibiting an inverse cascade, as expected for 2D turbulence, are electro-magnetically-driven flows \citep{kelley2011onset} and flows occurring on the surface of parametrically-excited waves\citep{von2011double, francois2014three, xia2017two}. These Faraday waves \citep{faraday1831xvii} have become a test-bed for the study of 2D-turbulence since the discovery of a horizontal surface flow, termed Faraday flow \citep{von2011double, von2013measurement, francois2013inverse}. However, early experiments were conducted in thin-layer systems, where the fluid depth is negligible compared to the characteristic length of the flow structures which are of about the size of the Faraday wavelength \(\lambda_F\). Recently, experiments at the fluid surface and at submerged planes in a non-shallow Faraday system characterized the sub-surface velocity fields \citep{colombi2021three}. The study unveiled the previously-unknown existence of 3D bulk flows with a rich variety of complex flow structures. An exponential decay in mean velocity magnitude and in mean vorticity with depth was observed, proving the confinement of the strong 2D turbulence to the fluid surface, as suggested by other theoretical and experimental work on wave-induced flows \citep{francois2014three, filatov2016nonlinear, levchenko2017faraday, francois2017wave}. The study \citep{colombi2021three} identified three regimes, namely the 2D Faraday flow at the surface, a transition regime of decay in velocity and vorticity right below the surface, and a bulk flow regime where large and slow three-dimensional structures are dominant. However, the mechanism of energy injection and transport through scales remained unclear: how are the surface and the bulk flow related in terms of energy and momentum transfer? What is the mechanism that fuels the fluid motion beneath the surface? Is an inverse energy cascade or at least a remnant of it also observed in the fluid bulk?
This study reports the co-existence of an inverse energy cascade, localized at the fluid surface, and a  direct  energy  cascade in  the  3D  bulk  flow  underneath in an experimental flow. Furthermore, sporadic vertical jets, which drive the vertical transport of energy and momentum from the surface to the bulk flow are found in the experimentally derived velocity fields.\\
Faraday waves are created by sinoidal vertical agitation on the surface of a layer of water (30~mm thickness) in a cylindrical container of 290~mm diameter as described previously in \citep{colombi2021three}. Time-dependent velocity fields at different horizontal and vertical planes are obtained by Particle Image Velocimetry (PIV) using a high-speed camera (Phantom Veo 640L) and either a rotatable laser light-sheet or blue LED lights. Details on the experimental set-up are presented in the supplemental material \citep{supplementalMaterial}. All experiments have been performed at a forcing frequency of 50~Hz and at two different acceleration amplitudes, $a_f=0.47$g and $a_f=0.70$g.
\begin{figure}[b]
\includegraphics[width=0.48\textwidth]{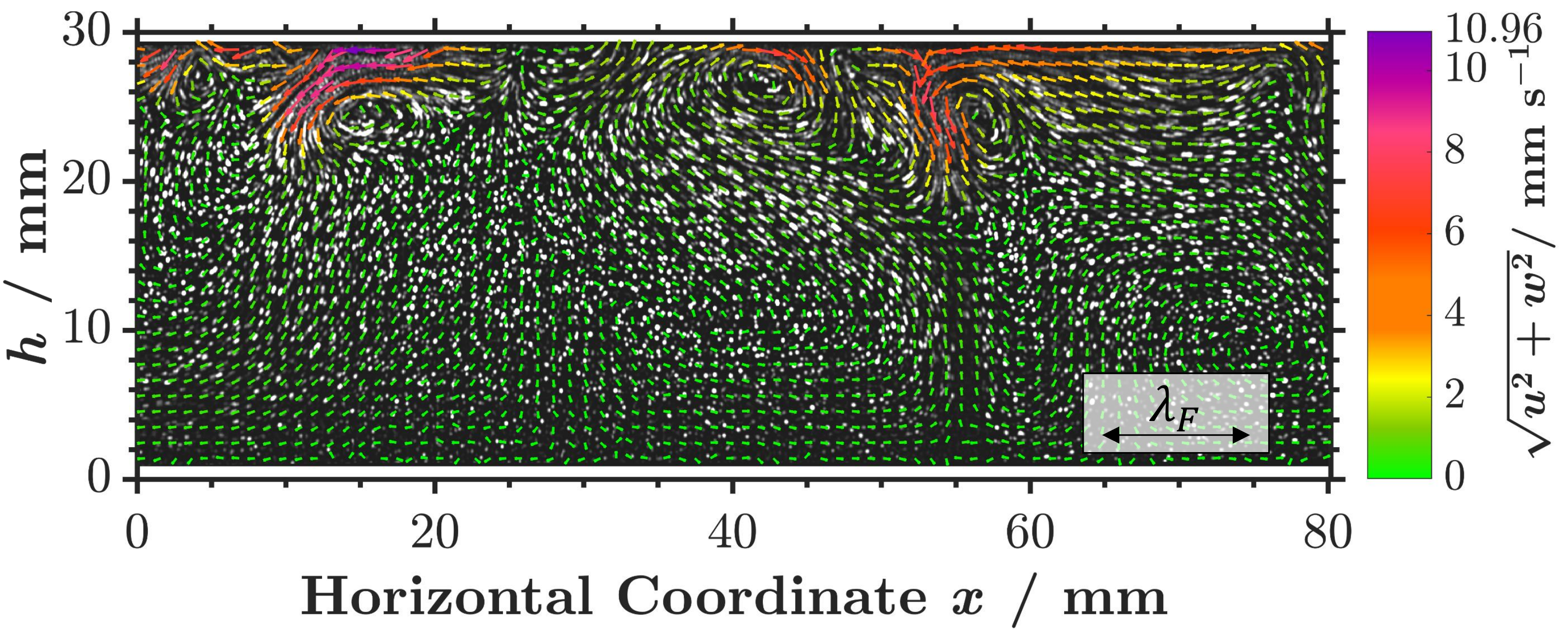}
\caption{\label{fig:vertical_jets} Instantaneous velocity field in the vertical plane for \(a_f=0.47\)g on top of the corresponding particle images averaged over 16 successive images (640~ms). Every second arrow shown. Color online.}
\end{figure} 
\noindent To analyse the turbulence at the surface and various submerged planes below it we calculate the energy and the enstrophy spectra in the frequency domain
. Beforehand, to reduce the spurious residual from the rectangular camera image, we smooth the spatial velocity and vorticity fields at the edges using a Tukey window function and zero-padding. The 2D energy and enstrophy spectra are subsequently averaged over all discrete time steps to improve statistical significance. Instantaneous and time-averaged spectra were found to yield similar results which suggests statistical convergence. The direction-independent energy spectra~$E(|\vec{k}|)$ are obtained by ring binning of the 2D energy fields \citep{singh2004energy}. To determine the inter-scale spectral flux through particular wavelengths~$\lambda_n,~n \in \mathbb{N}$, a spatial filter of varying width \citep{feldmann2020does, kelley2011spatiotemporal} originally used for large-eddy-simulations~(LES) \citep{alexakis2020local, natrajan2006role, vreman1994realizability}, is applied on the velocity fields. Low pass filtering is obtained by a smooth 4th-order Butterworth filter acting on each dimension separately in the frequency domain. The energy flux~$\Pi_E^{\lambda{_n}}=\widetilde{\tau_{ij}} \widetilde{s_{ij}},~i,~j \in {[1,2]}$ is calculated from the 2D data as described in \citep{liao2013spatial}, with \linebreak $\widetilde{s_{ij}} = \frac{1}{2} \bigl[\frac{\partial \widetilde{u}_i}{\partial x_j}+\frac{\partial \widetilde{u}_j}{\partial x_j}\bigr]$ being the strain rate of the filtered velocity fields and $\widetilde{\tau_{ij}} = \widetilde{u_i u_j} - \widetilde{u}_i \widetilde{u}_j$ the turbulent stress tensor. Analogously, an expression for the enstrophy flux, based on the vorticity \(\omega\), can be found \citep{kelley2011spatiotemporal}:
\begin{align}
\Pi_E^{\lambda{_n}} &=-\Bigl[ \widetilde{u_i u_j} - \widetilde{u}_i \widetilde{u}_j \Bigr] \frac{\partial \widetilde{u}_i}{\partial x_j}
\label{Energyflux}\\
\Pi_Z^{\lambda{_n}} &=-\Bigl[ \widetilde{u_i \omega} - \widetilde{u}_i \widetilde{\omega} \Bigr] \frac{\partial \widetilde{\omega}}{\partial x_i}
\label{Enstrophyflux}
\end{align}
A negative spectral flux $\Pi^{\lambda{_n}}<0$ implies a net energy transport towards smaller scales while a positive spectral flux $\Pi^{\lambda{_n}}>0$ a net energy transport towards larger scales. \\
In this study velocity measurements in the vertical plane enhance the understanding on the mechanisms of energy transport between the two-dimensional Faraday flow, confined at the fluid surface, and the intrinsically 3D flow that is observed underneath. For the following figures and diagrams, the velocity components in the three spatial directions are denoted as \(\smash{\textbf{\text{u}}=\left(u,v,w\right)^\top}\). Fig.~\ref{fig:vertical_jets} depicts an example of an instantaneous velocity field in the vertical plane for \(a_f=0.47\)g, highlighting the presence of strong vertical jets (at $x\approx 10$ and $x\approx 55$~mm) that cause an explosive downward transport of fluid from the surface to the bulk.  For both forcing accelerations the jets typically dissolve at about one Faraday wavelength below the surface. These jets are also observed in the horizontal velocity fields at the corresponding plane (\(h=21\)~mm) as fluid and momentum sources, i.e. positive divergence, with velocity arrows pointing outwards from a central point \citep{supplementalMaterial}. Because of the localized high velocity and strong vertical fluid transport, it is very reasonable to assume that these jets drive the flow structures below the surface and act as an energy injecting mechanism for the 3D bulk flow. Concurrently, their formation causes the generation of vortices on both sides of the jets with the vorticity pointing parallel to the liquid surface.\\
In Fig.~\ref{fig:vel_profiles} the sub-surface velocity profiles in dependence of the measurement height for both forcings, \(a_f=0.47\)g and \(a_f=0.70\)g (blue and red markers), are depicted. The velocities have been averaged over time and measurement runs additional to spatial averaging that has been carried out in the horizontal \(x\)- direction in the vertical measurements, and over all \(x\)- and \(y\)-locations in the horizontal planes. The diagram combines both vertical and horizontal PIV measurements, indicated with empty and filled markers respectively. Since in confined flows the velocity distributions are symmetric around zero, in particular for the highly-turbulent surface flow, the profiles are depicted for the absolute values of the velocity components. A good agreement is seen in the horizontal velocity components among the horizontal and vertical PIV measurements. The results highlight however different trends for the vertical and horizontal velocity components ($w$ and $u$). For both forcing accelerations, a strong exponential decay in \(u\) and \(v\) can be seen in a layer of approximately 10~mm (\(\approx\) one Faraday wavelength \(\lambda_F\)) below the surface, which is consistent with the results observed in \citep{colombi2021three}. At further depths, the slope of the decay decreases for both horizontal velocity components in a range between \(h=10\)-20~mm. The behaviour is somewhat different for the \(z\)-component of velocity, $w$, which at both forcing accelerations shows a steeper decay throughout the whole range between \(h=10\)-20~mm.  Due to the upper limits of the field of view, adjusted for dewarping and removal of reflections in the raw images, the full profile of the vertical velocity component can only be resolved up to \(h=28\)~mm such that no information about \(w\) is obtained directly at the air/water interface.  However, a slight tendency in \(w\) at \(h=28\)~mm to level out at a maximum value lower than the velocity in the horizontal plane can be recognized \vfill
\onecolumngrid 
\clearpage
\twocolumngrid
\noindent  (better observed on the linear-axis scale in the zoom-in region in Fig. \ref{fig:vel_profiles} b).  This is consistent  with the fact that there must be a zero vertical velocity boundary at the fluid surface. Towards the container bottom, in the third layer between \(h=0\) and 10~mm, all three velocity components show linear trends in rate of decline, with \(w\) approaching zero-velocity considerably earlier.\\
In order to further study the energy content in the bulk flow, the average ratio of flow kinetic energy in the vertical direction to the total flow kinetic energy is computed from the vertical PIV measurements: $\langle E_{\text{kin},z}\rangle/\langle E_{\text{kin,tot}}\rangle = \langle w^2\rangle / \langle u^2+v^2+w^2\rangle \approx \langle w^2\rangle /\langle u^2+u^2+w^2\rangle$.
\begin{figure}[t!]
\includegraphics[width=0.48\textwidth]{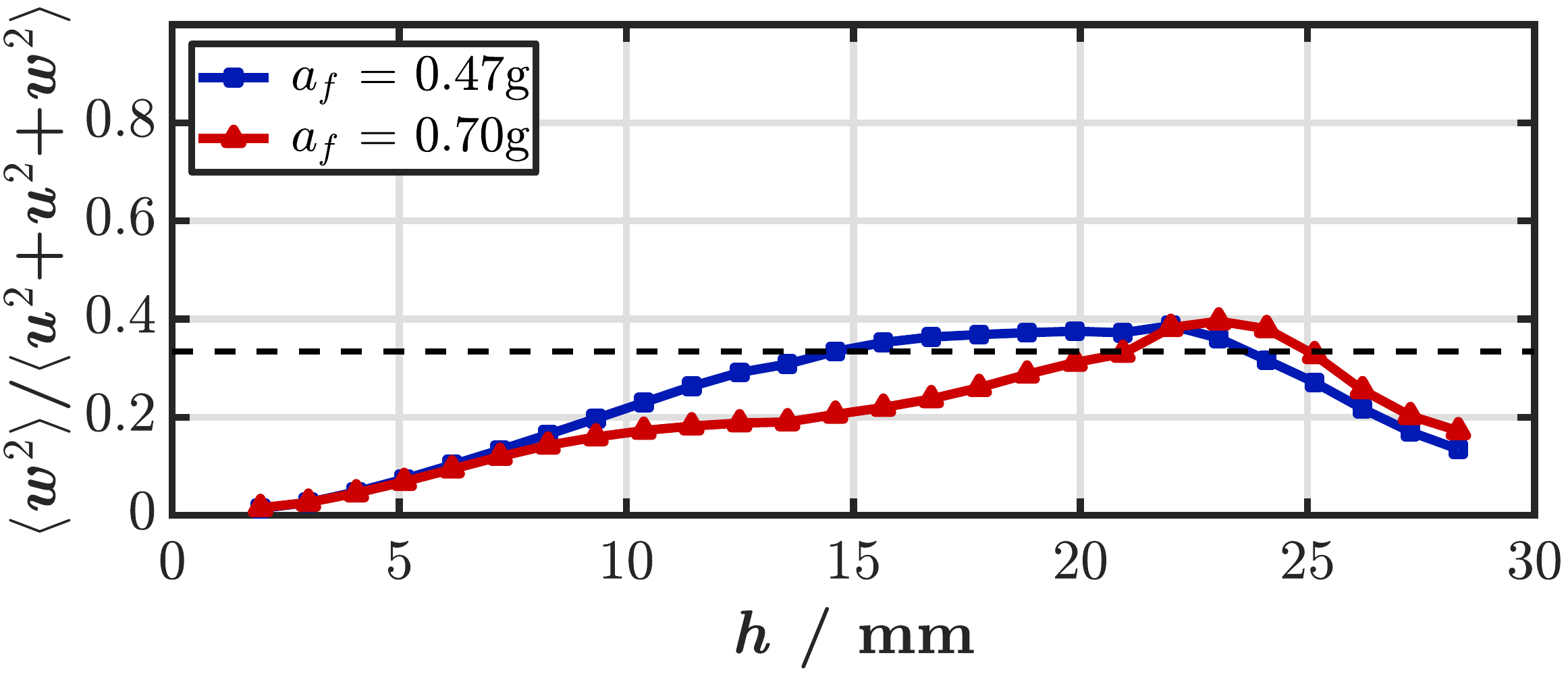}
\caption{\label{fig:vertical_energy} Average ratio of flow kinetic energy in the vertical direction to the total flow kinetic energy  along the height from the container bottom \(h\) for forcing accelerations \(a_f=0.47\)g and \(0.70\)g (blue squares and red circles). The horizontal dashed black line shows the 3D isotropic turbulence case at value 1/3. Color online.}
\end{figure}
\noindent Thereby we assume that \(\langle u^2\rangle \approx \langle v^2 \rangle\) is valid on a statistical average over time and space, as also verified by the very similar profiles of \( \langle|u|\rangle\) and \( \langle|v|\rangle\) in Fig.~\ref{fig:vel_profiles}. The average ratio of the flow kinetic energy profile along the height \(h\) is depicted in Fig.~\ref{fig:vertical_energy}. The vertical component of the kinetic energy gradually increases to peak in a layer of approximately one Faraday wavelength below the surface. In this layer, the energy content in the \(z\)-direction becomes dominant (larger than 1/3). This can be explained by the presence of the confined and strong downwards jets seen in the velocity fields (Fig.~\ref{fig:vertical_jets}). At lower depths, for \(h<20\)~mm, the influence of $\langle w^2\rangle$ gradually decreases and becomes negligibly small close to the container bottom (around 5\%).
In summary, the energy content below Faraday waves is dominated by horizontal flow components but a local peak emerges where the vertical motion caused by intense jets becomes dominant. 
From the analysis of velocity structures and kinetic energy profiles, four horizontal planes are determined for the spectral analysis of energy transport and net energy and enstrophy fluxes. One plane is located at the surface, \(h=30\)~mm,  where we expect to see features of 2D turbulence in the Faraday flow. A second submerged plane is placed immediately below at \(h=27\)~mm, in the layer where we observed the vertical jets and concurrent vorticity pointing parallel to the fluid surface. The third plane is placed at \(h=21\)~mm, the depth where jets typically dissolve and where most energy is contained in vertical motion. For completeness, a last measurement plane is located at \(h=4\)~mm, in the vicinity of the container bottom (not shown). The results are depicted in Fig.~\ref{fig:energy_spectra} for \(h=30\), 27 and 21~mm and both forcing accelerations (note the different scales among the panels). The results in panel Fig.~\ref{fig:energy_spectra}~a clearly validate the double-cascade feature of 2D turbulence in the Faraday flow on the fluid surface \citep{von2011double, von2013measurement, francois2013inverse}. For both forcing accelerations, it is possible to recognize a sharp bend in the energy spectrum, occurring at about the energy injection wavenumber \(k_{\text{inj}} \approx 2\pi/\lambda_F = 1.25\)~mm\(^{-1}\) (computed with the Faraday wavelength as the dominant forcing scale for energy injection). \vfill
\onecolumngrid
 \clearpage
 \twocolumngrid
  \noindent  At wavenumbers \(k<k_{\text{inj}}\) the slope of \(k^{-5/3}\) is captured fairly well, and the negative net energy flux \(\Pi_E(k)\) validates the presence of an inverse energy cascade, although the transition from negative to positive is not localized at one wavenumber as in theory or simulated flows \citep{chen2006physical, biferale2017two} but occurs more gradually. Our data also resolves the direct enstrophy cascade with positive net enstrophy flux \(\Pi_Z(k)\) for \(k>k_{\text{inj}}\). The slope is however slightly steeper than the \(k^{-3}\) scaling predicted by Kraichnan \citep{kraichnan1967inertial}.
This phenomenon is not uncommon in experimental 2D turbulence, where friction and damping effects can cause deviations at larger wavenumbers in contrast to theory or simulated data \citep{francois2013inverse, chen2006physical}.
The situation is substantially different in the plane immediately below the fluid surface, at \(h=27\)~mm, shown in panel Fig.~\ref{fig:energy_spectra}~b. The slope of the energy spectra is more homogeneous throughout the entire wavenumber range and no distinct bend can be seen at the energy injection wavenumber. For the stronger forcing acceleration \(a_f=0.70\)g, the slope approaches the -5/3 value expected for both, the direct 3D and inverse 2D cascade. Here, in sharp contrast to the surface flow, the net energy flux remains positive for all wavenumbers.
 This fact proves that, here, the flow exhibits a direct 3D cascade and that there is a transition from the 2D turbulent Faraday flow on the surface to a direct energy cascade and a 3D flow in the bulk. This observation is validated by the zero net enstrophy flux, indicating a vanishing influence of the direct enstrophy cascade. The results add to the results on 2D3C flows as described in \citep{biferale2017two, kokot2017active} by proving the simultaneous existence of a 2D and a 3D cascade in an experimental flow. Both fundamentally different transport mechanisms for energy and momentum are clearly found, and the flows interact in close proximity to the fluid surface. A possible pathway of the energy could therefore be as follows: In a thin surface layer at the air/water interface, energy is transferred upscale by the inverse cascade until the surface energy spectra tend to level out at wavenumbers \(k\lesssim 0.2\) (Fig.~\ref{fig:energy_spectra}~a,~b). There, energy might be lost to structures that cause the sporadic downward jets thereby fueling the direct cascade found in the bulk flow below.  
The results for the energy fluxes and spectra are similar at further submerged planes below the surface (e.g. \(h=21\)~mm, Fig.~\ref{fig:energy_spectra}~c), where both trends of positive net energy flux and zero net enstrophy cascade are confirmed, although with much smaller magnitudes than at \(h=27\)~mm. This is also observed in the velocity fields, as the flow becomes less turbulent, and the velocity structures larger, less chaotic and with longer temporal scales \citep{colombi2021three} and \citep{supplementalMaterial}.\\
To conclude, we have performed the analysis of energy spectra and mean spectral energy fluxes in non-shallow Faraday waves, obtained from planar PIV measurements at different submerged planes and the surface. For the first time a 2D inverse energy cascade on the fluid surface and a co-existing 3D direct energy cascade in the bulk flow underneath were observed in an experimental flow. This observation adds to the existing research on the interplay between two- and three-dimensional flows \citep{kelley2011onset, francois2014three, francois2017wave, biferale2017two, kokot2017active}. The double cascade regime of the Faraday flow is well captured by our experiments, clearly showing a change in the energy spectral slope at about the energy injection wavenumber \(k_{\text{inj}}=1.25\)~mm\(^{-1}\) and an inverse energy cascade with negative net energy flux towards larger scales. At further depths the transition to 3D turbulence becomes evident due to the exclusively positive net energy flux and zero enstrophy flux throughout the entire wavenumber range. Additionally, measurements of velocities \( (u,w) \) in the container's vertical cross-section unveiled the presence of strong and confined vertical jets. These originate from the surface and dissolve at approximately one Faraday wavelength (\(\approx 10\) mm) below it. The jets, together with the simultaneous formation of vortices by shear effects, appear to be the main fuel for the three-dimensional bulk flow beneath Faraday waves. Our results further reveal that the vertical component of velocity decreases at a smaller rate than the horizontal components in the aforementioned transition layer, directly beneath the fluid surface. Conversely, we find that the average ratio of the flow kinetic energy in the \(z\)-direction to the total kinetic energy increases in this layer, indicating a gradual shift from the 2D Faraday flow confined to the fluid surface towards a 3D bulk flow.\\
\noindent \textbf{Funding}
The authors gratefully acknowledge the financial support provided by the Deutsche
 Forschungsgemeinschaft (DFG) within the project 395843083 (KA 4854/1-1).
\bibliographystyle{apsrev4-2}
\bibliography{short_journal_names,PRL_Transition_2021_corrections.bib}

\begin{thebibliography}{23}
\bibitem [{sup()}]{supplementalMaterial}%
  \BibitemOpen
  \href@noop {} {\bibinfo {title} {See {S}upplemental {M}aterial at [url will
  be inserted by publisher] for description of experimental and analytical
  methods and additional flowfield figures.}}\BibitemShut {Stop}%
\end{thebibliography}

\end{document}